\documentclass[prl,twocolumn,twoside,showpacs,floatfix,superscriptaddress]{revtex4}

\usepackage{graphicx,color,rotating}
\usepackage{amsmath,amssymb}
\usepackage{dcolumn}
\usepackage{times,txfonts}
\usepackage{pifont}
\usepackage{dsfont}

\newcommand{\elem}[2]{$^{\text{#2}}$#1}
\newcommand{\eq}[1]{\begin{equation}#1\end{equation}}

\newcommand{\bra}[1]{\ensuremath{\langle{#1}|\,}}

\newcommand{\ket}[1]{\ensuremath{\,|{#1}\rangle}}

\newcommand{\op}[1]{\ensuremath{\hat{#1}}}

\newcommand{\linemediumsolid}[1][black]{\unitlength0.5ex
  {\color{#1}
  \begin{picture}(6,1)
  \linethickness{0.4mm}
  \put(0,1){\line(1,0){6.0}}
  \end{picture}}\nolinebreak
}

\newcommand{\linemediumdashed}[1][black]{\unitlength0.5ex
  {\color{#1}
  \begin{picture}(6,1)
  \linethickness{0.4mm}
  \put(0,1){\line(1,0){1.5}}
  \put(2.2,1){\line(1,0){1.5}}
  \put(4.4,1){\line(1,0){1.5}}
  \end{picture}}\nolinebreak
}

\newcommand{\clebsch}[6]{ \ensuremath{\!\left(\!\!
\begin{array}{cc}
 {#1} & \!\!\!\!{#2} \\
 {#4} & \!\!\!\!{#5} 
\end{array}
 \!\!\right|
\left.\!\!\!
\begin{array}{c}
 {#3}\\
 {#6}
\end{array}\!\! \right)\!}
}

\begin{document}

\title{Continuum and Three-Nucleon Force Effects on \elem{Be}{9} Energy Levels}

\author{Joachim Langhammer}
\email{joachim.langhammer@physik.tu-darmstadt.de}
\affiliation{Institut f\"ur Kernphysik, Technische Universit\"at Darmstadt, 64289 Darmstadt, Germany}

\author{Petr Navr\'atil}
\email{navratil@triumf.ca}
\affiliation{TRIUMF, 4004 Wesbrook Mall, Vancouver, British Columbia, V6T 2A3, Canada}

\author{Sofia Quaglioni}
\affiliation{Lawrence Livermore National Laboratory, P.O. Box 808, L-414, Livermore, CA 94551, USA}

\author{Guillaume Hupin}
\altaffiliation{Present address: Department of Physics, University of Notre Dame, Notre Dame, IN 46556, USA}
\affiliation{Lawrence Livermore National Laboratory, P.O. Box 808, L-414, Livermore, CA 94551, USA}

\author{Angelo Calci}
\affiliation{Institut f\"ur Kernphysik, Technische Universit\"at Darmstadt, 64289 Darmstadt, Germany}
\affiliation{TRIUMF, 4004 Wesbrook Mall, Vancouver, British Columbia, V6T 2A3, Canada}

\author{Robert Roth}
\email{robert.roth@physik.tu-darmstadt.de}
\affiliation{Institut f\"ur Kernphysik, Technische Universit\"at Darmstadt, 64289 Darmstadt, Germany}

\date{\today}

\begin{abstract}  
We extend the recently proposed ab initio no-core shell model with continuum to include three-nucleon (3N) interactions beyond the few-body domain. The extended approach allows for the assessment of effects of continuum degrees of freedom as well as of the 3N force in ab initio calculations of structure and reaction observables of $p$- and lower-$sd$-shell nuclei. As first application we concentrate on energy levels of the \elem{Be}{9} system for which all excited states lie above the n-\elem{Be}{8} threshold. For all energy levels, the inclusion of the continuum significantly improves the agreement with experiment, which was an issue in standard no-core shell model calculations. Furthermore, we find the proper treatment of the continuum indispensable for reliable statements about the quality of the adopted 3N interaction from chiral effective field theory. In particular, we find the $\frac{1}{2}^+$ resonance energy, which is of astrophysical interest, in good agreement with experiment.
\end{abstract}


\maketitle

\paragraph{Introduction.} Over recent years the inclusion of three-nucleon (3N) interactions into different ab initio approaches for nuclear structure calculations has been challenging but successfully completed with a variety of interesting applications~\cite{BarNa13,RoLa11,RoBi12,HaPa13,HaHj12b,BiLa13, BiLa14,CiBa13,SoCi14,HeBo13,HeBi13,EpKr12,EpKr14,LaEp14,BoHe14,JaEn14}. However, beyond the few-body domain, the inclusion of 3N interactions in ab initio studies of continuum effects in weakly-bound systems or nuclear reactions have been completed for the five-nucleon system only, e.g., within the Green's function Monte-Carlo (GFMC) approach~\cite{NoPi07} and the no-core shell model combined with the resonating group method (NCSM/RGM)~\cite{HuLa13, QuNa09}.
To arrive at a more efficient unified ab initio theory applicable to nuclear structure and reactions on equal footing, the NCSM/RGM approach has been recently generalized to the no-core shell model with continuum (NCSMC)~\cite{BaNa13b,BaNa13}. In Ref.~\cite{HuQu14}, the NCSMC was applied to p-$^4$He scattering for the first time including 3N interactions using an algorithm restricted to $A{=}3$ and $A{=}4$ target nuclei.
In this communication, we extend the NCSMC formalism to include 3N interactions in a general framework applicable to arbitrary $p$- and lower-$sd$ shell target nuclei. This is a major step towards a refined nuclear structure and reaction theory that allows ab initio studies of observables affected by continuum degrees of freedom starting from the best Hamiltonians currently available. This is vital to provide robust QCD-based predictions starting from chiral effective field theory interactions, e.g., for light exotic nuclei.

As a first application, we study the effects of the continuum and of the 3N interaction on the energy levels of the \elem{Be}{9} nucleus.
This system is interesting because only its ground-state is bound, while all excited states are unstable and subject to neutron emission as the n-\elem{Be}{8} threshold energy is located experimentally at $1.665\,\text{MeV}$~\cite{TiKe04}, being the lowest neutron threshold of all stable nuclei. Therefore, it is appealing to study the impact of the continuum on the excited-state resonances with particular focus on the effects of the chiral 3N interactions. Earlier studies of these energy levels within the NCSM showed problems with the model-space convergence, and, in particular the positive-parity states were found too high in excitation energy compared to experiment~\cite{FoNa05, BuNe10}. Also the splitting between the lowest $5/2^-$ and $1/2^-$ states is found overestimated in no-core shell model (NCSM) calculations using the INOY interaction model that includes 3N effects~\cite{FoNa05}. Moreover, GFMC calculations~\cite{PiVa02} show strong sensitivity of the splitting with respect to 3N interactions. As including the 3N effects appeared to shift the splitting away from the experiment, these studies seemed to highlight deficiencies of 3N force models. 
Furthermore, \elem{Be}{9} is interesting for astrophysics, because it provides seed material for the production of \elem{C}{12} in the explosive nucleosynthesis of core-collapse supernovae via the ($\alpha\alpha$n)\elem{Be}{9}($\alpha$n)\elem{C}{12} reaction, an alternative to the triple-$\alpha$ reaction~\cite{EfNe14, SaKa05, BuNe10} bridging the $A=8$ instability gap and triggering the $r$ process. In particular, the description of the first $\frac{1}{2}^+$ state of \elem{Be}{9} slightly above threshold poses a long standing problem~\cite{KuRi87,BuNe10,EfNe14}, relevant for the cross sections and reaction rates.

\paragraph{The No-Core Shell Model with Continuum and 3N forces.}

To arrive at the ab initio description of the \elem{Be}{9} nucleus we generalize the NCSMC~\cite{BaNa13b,BaNa13} to include 3N interactions. In the following we highlight the quantities  affected by the inclusion of 3N interactions, while we refer to Ref.~\cite{BaNa13} for details about the general formalism and the implementation of the NCSMC. In the NCSMC the eigenstates of the $A$-body system are represented by
\eq{
\ket{\Psi_{A}^{J\pi T}} = \sum_{\lambda} c_{\lambda} \ket{A\lambda J^{\pi} T} + \sum_{\nu} \int \mathrm{d}r r^2 \frac{\gamma_{\nu}(r)}{r} \op{\mathcal{A}}_{\nu} \ket{\Phi_{\nu r}^{J\pi T}}\,,
\label{eq:NCSMC-expansion}
}
constituting an expansion in an over-complete basis of non-orthogonal states.
The first term consists of a superposition of eigenstates of the $A$-nucleon system computed within the NCSM. These describe the correlations between all nucleons in localized configurations, but have limitations concerning the description of nuclear clustering and scattering states. This is cured by the expansion in $(A{-}a,a)$ binary-cluster channel states, $\ket{\Phi^{J\pi T}_{\nu r}}$, in the second term of Eq.~\eqref{eq:NCSMC-expansion}. They are defined as
\eq{
\ket{\Phi^{J\pi T}_{\nu r}} = \Big[\big(\ket{A-a\, \alpha_1 I_1^{\pi_1} T_1}\ket{a\, \alpha_2 I_2^{\pi_2} T_2}\big)^{(sT)} \ket{rl}\Big]^{(J\pi T)}\,,
}
with the collective index $\nu=\{A{-}a, \alpha_1, I_1^{\pi_1}, T_1,a, \alpha_2, I_2^{\pi_2}, T_2,$ $ s l\}$, and the state \ket{r l} describing the
relative motion of the clusters with relative distance $r$ and relative orbital angular momentum $l$~\cite{QuNa09}. The expansion coefficients $c_{\lambda}$ and the continuous relative-motion amplitudes $\gamma_{\nu}(r)$ result from the Schr\"odinger equation,
\eq{
\begin{pmatrix}
H_{\text{NCSM}}&\bar{h}\\
\bar{h}                         & \bar{\mathcal{H}}
\end{pmatrix}
\begin{pmatrix}
c\\\, \chi\,
\end{pmatrix}
=
E \begin{pmatrix}
\mathds{1} & \bar{g}\\
\bar{g}                & \mathds{1}
\end{pmatrix}
\begin{pmatrix}
c\\\,\chi\,
\end{pmatrix}\,,
\label{eq:NCSMC-Equation-orth}
}
represented using expansion~\eqref{eq:NCSMC-expansion}.
Here orthogonalized channel states have been used, yielding the relative-motion wave functions $\chi(r)$ as detailed in Refs.~\cite{BaNa13,QuNa09}. The generalization of the NCSMC formalism to 3N interactions affects the Hamiltonian contributions in Eq.~\eqref{eq:NCSMC-Equation-orth}. First of all, the $A$- and $(A-a)$-body NCSM eigenstates are computed including the 3N interaction. The former affect the eigenvalues that enter the diagonal matrix $H_{\text{NCSM}}$ describing the NCSM sector of the Hamiltonian kernel. Additionally, in the NCSM/RGM Hamiltonian kernel $\bar{\mathcal{H}}$ new contributions need to be considered, i.e., for the present $a{=}1$ case three-body densities generated by the 3N interaction as discussed in Ref.~\cite{HuLa13}. Here, we rely on the formalism using uncoupled densities to access $A\geq5$ nuclei. Finally, we generalize the Hamiltonian form factor $\bar{h}$ to include 3N forces. 
Again, we adapt the algorithm with uncoupled densities introduced in Ref.~\cite{HuLa13} and obtain for the 3N interaction contribution to the Hamiltonian form factor before orthogonalization
\begin{align}
_{\rm SD}\langle&A \lambda J^\pi M_J T M_T\big| \sqrt{A} \tfrac{1}{2}(A-1)(A-2)  V^{3N}_{A-2\,A-1\,A} \ket{\Phi^{J^\pi T}_{\kappa}}_{\rm SD} \notag \\
= &\frac{1}{12}\sum_{M_1 m_j} \sum_{M_{T_1} m_t} 
\, \clebsch{I_1}{j}{J}{M_1}{m_j}{M_J} \; \clebsch{T_1}{\tfrac{1}{2}}{T}{M_{T_1}}{m_t}{M_T} 
\; \sum_{abcef}\bra{cba}V^{3N}\ket{fe \tilde{j}\:} 
\notag \\
  _{\rm SD}&{\bra{A \lambda J^\pi M_J T M_T}} a^{\dagger}_{a} a^{\dagger}_{b} a^{\dagger}_{c} 
a_{f} a_{e} \ket{ A{-}1\, \alpha_1 I_1^{\pi_1} M_1 T_1 M_{T_1}}_{\rm SD} \, , \label{NNN-h-express}
\end{align}
with
\begin{align}
|\Phi^{J^\pi T}_{\kappa}\rangle_{\rm SD} &= \Big[\left|A{-}1\, \alpha_1 I_1^{\pi_1} T_1\right\rangle_{\rm SD} |n \ell j \tfrac12\rangle \Big]^{(J^\pi T)} \notag \\
&= \sum_{M_1 m_j} \sum_{M_{T_1} m_t} \; \clebsch{I_1}{j}{J}{M_{1}}{m_j}{M_J} \;\;\clebsch{T_1}{\tfrac{1}{2}}{T}{M_{T_1}}{m_t}{M_T}\notag \\
&\times |A{-}1\, \alpha_1 I_1^{\pi_1} M_1 T_1 M_{T_1}\rangle_{\rm{SD}} |n\ell jm_j\tfrac{1}{2}m_t\rangle \label{eq:new-basis-formula}
\end{align}
where $|\tilde{j}\rangle{\equiv} |n\ell jm_j\tfrac{1}{2}m_t\rangle$ denotes the HO single-particle state of the projectile and the subscript SD denotes Slater determinant channel states~\cite{QuNa09}. To solve the NCSMC equations~\eqref{eq:NCSMC-Equation-orth} we use the coupled-channel $R$-matrix method on a Langrange mesh~\cite{DeBa10, HeRo02, Ba02}. We adopt a channel radius of $18\,\text{fm}$ and $40$ mesh points throughout, and have checked that our calculations are independent of these parameters.

\paragraph{SRG-Transformed NN+3N Interactions.}

We use the nuclear interactions derived from chiral effective field theory~\cite{EpHa09, MaEn11}, namely the NN interaction at N$^3$LO by Entem \& Machleidt~\cite{EnMa03} and the 3N interaction at N$^2$LO in its local form~\cite{Na07} with cutoff $\Lambda_{\text{3N}} = 400\,\text{MeV}/c$~\cite{RoBi12}. This choice is motivated by the observation that the $\Lambda_{\text{3N}} = 500\,\text{MeV}/c$ Hamiltonian overbinds the n-\elem{Be}{8} threshold by about 800 keV in importance-truncated NCSM (IT-NCSM) calculations at $N_{\text{max}} = 12$, unlike the reduced-cutoff 3N interaction. Additionally, we soften the interactions, using the similarity renormalization group (SRG)\cite{SzPe00,BoFu07,JuNa09,RoLa11,RoCa14}. Depending on whether we include the chiral 3N interaction in the initial Hamiltonian or not, we arrive at the NN+3N-full and the NN+3N-induced Hamiltonians, respectively. Both Hamiltonians include SRG-induced 3N interactions but neglect SRG-induced four- and multi-nucleon interactions. However, the latter are typically negligible for nuclei with less than 10 nucleons as discussed in Refs.~\cite{RoLa11, JuNa09, RoCa14}. 


\begin{figure}[t]
\includegraphics[width=1\columnwidth]{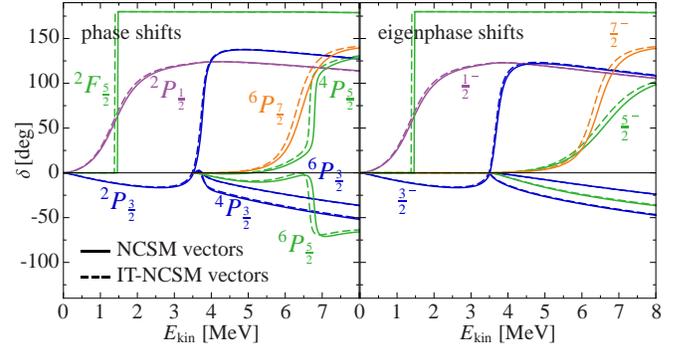}
\caption{(color online)
NCSMC n-\elem{Be}{8} phase shifts and eigenphase shifts for negative parity at $N_{\text{max}}=8$ computed with NCSM eigenvectors (\linemediumsolid) compared to the use of IT-NCSM vectors (\linemediumdashed). Remaining parameters are $\hbar\Omega=20\,\text{MeV}$, $\alpha=0.0625\,\text{fm}^4$, and $E_{\text{3max}}=14$. Same colors correspond to identical angular momenta.
}
\label{fig:IT_check}
\end{figure}

\paragraph{(Eigen)Phase Shifts.}

As first application of the NCSMC with explicit 3N interactions we focus on the excitation spectrum of \elem{Be}{9} up to about $8\,\text{MeV}$ above the n-\elem{Be}{8} threshold. We include in the first term of expansion~\eqref{eq:NCSMC-expansion} all \elem{Be}{9} states we find in this energy range with the NCSM. These are the four positive-parity states $\frac{1}{2}^+$, $\frac{5}{2}^+$, $\frac{3}{2}^+$, $\frac{9}{2}^+$, and the six negative-parity states $\frac{3}{2}^-$, $\frac{5}{2}^-$, $\frac{1}{2}^-$, $\frac{3}{2}^-$, $\frac{7}{2}^-$, and $\frac{5}{2}^-$. This selection is consistent with experimental data~\cite{TiKe04} showing a gap of about $3\,\text{MeV}$ between the second $\frac{5}{2}^-$ state we include, and the next known resonance at $11.2\,\text{MeV}$, which is also found near this energy in the NCSM. For the second term of expansion~\eqref{eq:NCSMC-expansion} we restrict ourselves to channels with single-neutron projectiles and \elem{Be}{8} targets. For \elem{Be}{8} we include the $0^+$ ground as well as its first excited $2^+$ state from the NCSM, i.e., in  a bound-state approximation.

\begin{figure}[t]
\includegraphics[width=1\columnwidth]{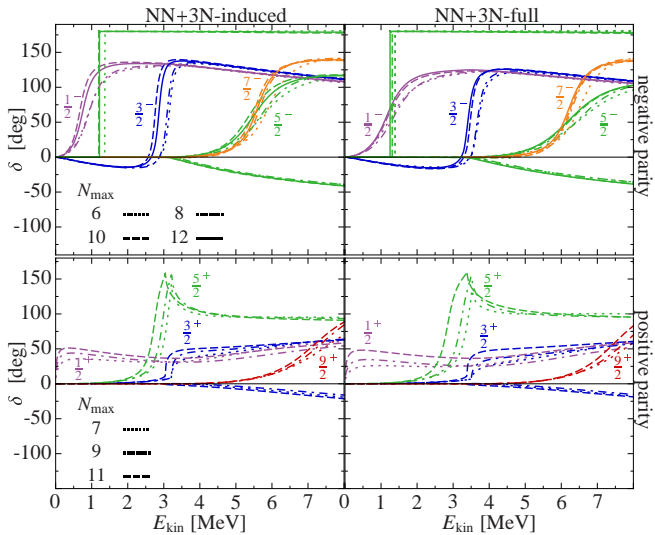}
\caption{(color online) $N_{\text{max}}$ dependence of NCSMC n-\elem{Be}{8} eigenphase shifts. The left- and right-hand columns show the results for the NN+3N-induced and NN+3N-full Hamiltonian, respectively. Remaining parameters identical to Fig.~\ref{fig:IT_check}.
}
\label{fig:NmaxDep}
\end{figure}

Evidently, the NCSMC relies on a set of NCSM eigenstates whose computation quickly becomes demanding for large model spaces. To cope with this we use the IT-NCSM. This reduces the computational cost not only for the input NCSM vectors, but, even more importantly, also for the NCSM/RGM kernels $\bar{\mathcal{H}}$ and the NCSMC coupling form factors $\bar{g}$ and $\bar{h}$, because only relevant Slater determinants are considered in the expansion of the computed eigenstates. As discussed in Ref.~\cite{Roth09} the IT-NCSM usually includes the a posteriori extrapolation to the full NCSM space. For (eigen)phase shifts we waive such extrapolations and use IT-NCSM eigenvectors computed with the smallest importance thresholds $\kappa_{\text{min}}=2\cdot10^{-5}$ and $C_{\text{min}}=10^{-4}$. We assess the quality of the IT in Fig.~\ref{fig:IT_check} by direct comparison with phase shifts computed with full NCSM vectors in the largest feasible model space, i.e., $N_{\text{max}}=8$. Overall, we find the (eigen)phase shifts with and without the use of the IT on top of each other. The only exceptions are the $\frac{5}{2}^-$ and $\frac{7}{2}^-$ resonance positions near $7\,\text{MeV}$ for which we find differences of about $100\,\text{keV}$, while the non-resonant $\frac{5}{2}^{-}$ eigenphase shift is not affected. For the results presented in the following we use IT-NCSM eigenvectors  for $N_{\text{max}}\geq7$. Furthermore, we concentrate the following discussions on the physically more relevant eigenphase shifts, while phase-shift plots are included in the Supplemental Material~\cite{Supp}.

Next, we analyze the convergence of the n-\elem{Be}{8} eigenphase shifts with respect to the HO model-space parameter $N_{\text{max}}$. In Fig.~\ref{fig:NmaxDep} we show the results for the NN+3N-induced and NN+3N-full Hamiltonians. For negative parity changing the model-space size almost exclusively affects the resonance positions. The typical convergence pattern shows a shift of the resonance positions when going from $N_{\text{max}}=8$ to $10$ but only a minor change from the step to $N_{\text{max}}=12$. Exceptions are the $\frac{7}{2}^-$ and the first $\frac{5}{2}^-$ resonances that are practically independent of $N_{\text{max}}$. We do not expect that an $N_{\text{max}}=14$ calculation would significantly change the present results. Also the positive-parity eigenphase shifts are most sensitive to the model-space size near resonances. The sole exception is the $\frac{1}{2}^{+}$ eigenphase shift that is affected at all energies. Overall, the $N_{\text{max}}$ dependence is stronger than for the negative-parity partial waves. Nevertheless, we use the $N_{\text{max}}=11$ results for the investigation of the positive-parity spectrum of \elem{Be}{9} in the following.

\begin{figure}[t]
\includegraphics[width=1\columnwidth]{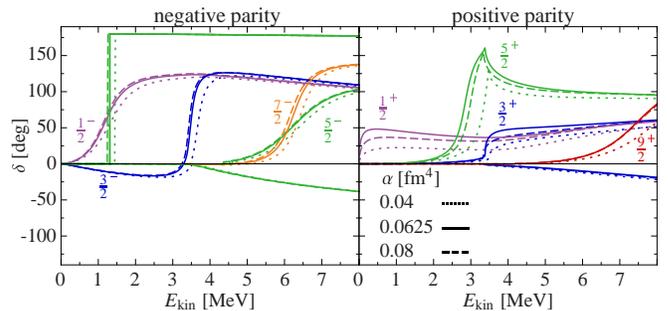}
\caption{(color online) 
SRG flow-parameter dependence of the NCSMC n-8Be eigenphase shifts for the NN+3N-full Hamiltonian at $N_{\text{max}}=10\,(11)$ for negative (positive) parity. Remaining parameters identical to Fig.~\ref{fig:IT_check}.
}
\label{fig:alphaDep}
\end{figure}

Since we use SRG-transformed Hamiltonians we check the dependence of our results on the flow parameter $\alpha$. Although flow-parameter dependencies are typically negligible in the domain of light nuclei~\cite{RoLa11, RoCa14} and we employ the 3N interaction with cutoff $\Lambda_{\text{3N}}=400\,\text{MeV}/c$ that additionally reduces SRG-induced multi-nucleon forces~\cite{RoBi12}, possible non-convergences or inconsistent truncations may nevertheless cause $\alpha$ dependencies, cf.~Refs.~\cite{HuLa13, BiLa13,BiLa14,HeBo13}. Hence, we study the n-\elem{Be}{8} eigenphase shifts for flow parameters $\alpha=0.04$, $0.0625$, and $0.08\,\text{fm}^4$ for the NN+3N-full Hamiltonian in Fig.~\ref{fig:alphaDep}. For negative parity we find only negligible differences between the eigenphase shifts for $\alpha=0.0625$ and $0.08\,\text{fm}^4$. In contrast, for $\alpha=0.04\,\text{fm}^4$ we observe larger deviations at least near resonances. This is most likely due to the slower convergence of the many-body calculation for this smaller flow-parameter as the direction of the deviations is consistent with the $N_{\text{max}}$-convergence pattern of Fig.~\ref{fig:NmaxDep}. For positive parity the overall conclusions are identical, however, the differences are larger compared to the negative-parity results. Again this may be tied to the slower rate of convergence we observed for positive parity in Fig.~\ref{fig:NmaxDep}. For these reasons we use $\alpha=0.0625\,\text{fm}^4$ in the following.

As argued above, contributions to the $\alpha$ dependence could also originate from additional truncations. 
Although we exploit the $JT$-coupled storage scheme for the 3N matrix elements~\cite{RoLa11, RoCa14} we have to truncate the set of 3N matrix elements by specifying a maximum three-nucleon energy via $E_{\text{3max}}\leq e_1+e_2+e_3$ with $e_i$ as single-particle HO energy quantum number. For all calculations presented here we use $E_{\text{3max}}=14$. For the NCSM/RGM kernels $\bar{\mathcal{H}}$ we found almost no dependence on $E_{\text{3max}}$ in calculations of neutron elastic scattering on \elem{He}{4}~\cite{HuLa13}. The sensitivity of the NCSMC Hamiltonian form factors $\bar{h}$ is still smaller, because of the $N_{\rm max}$ truncation of the composite eigenstates.

\paragraph{\elem{Be}{9} Energy Levels.}

\begin{figure}[t]
\includegraphics[width=0.85\columnwidth]{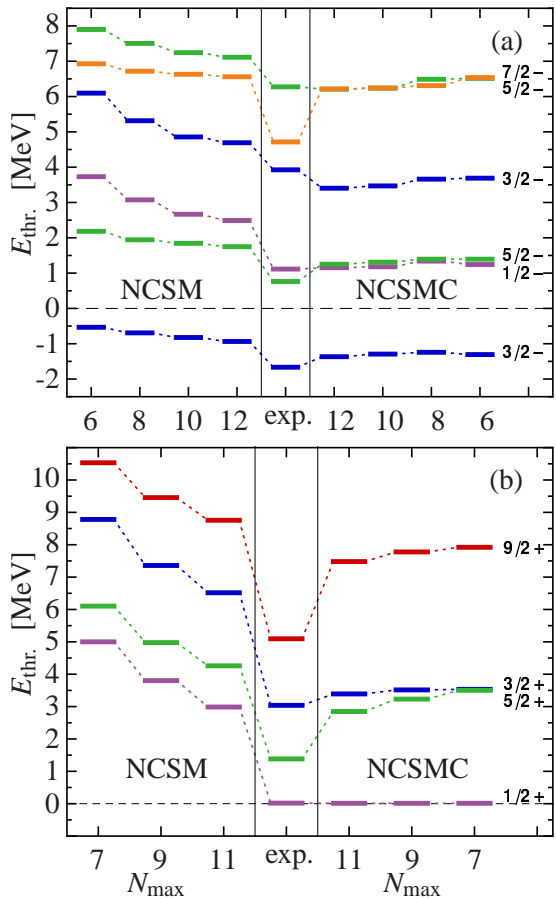}
\caption{(color online) 
Negative~(a) and positive~(b) parity spectrum of \elem{Be}{9} relative to the n-\elem{Be}{8} threshold  as function of $N_{\text{max}}$ for the NCSM (left-hand columns) and NCSMC (right-hand columns) compared to experiment~\cite{TiKe04}. Remaining parameters are $\hbar\Omega = 20\,\text{MeV}$ and $\alpha=0.0625\,\text{fm}^4$. See text for further explanations.
}
\label{fig:Spectra_Nmax}
\end{figure}

For a direct assessment of the impact of the continuum and the 3N interactions on the \elem{Be}{9} energy levels, we extract the resonance centroids $E_R$ with corresponding widths $\Gamma$ from the eigenphase shifts shown in Fig~\ref{fig:NmaxDep}. Following Refs.~\cite{BaNa13,BaNa13b}, we extract the centroids from the maximum of the derivative of the eigenphase shifts with respect to the kinetic energy, i.e., $E_R$ is defined by the inflection point of the eigenphase shifts. The width follows from
$\Gamma=2/\big(\mathrm{d\delta(E_{\text{kin}})}/\mathrm{d}E_{\text{kin}} \big)\big|_{E_{\text{kin}}=E_R}$
with eigenphase shifts $\delta$ in units of radians~\cite{ThNu09}.
Besides resonances, the NCSMC approach also yields information about bound states if we apply the $R$-matrix approach with bound-state boundary conditions~\cite{BaHe98}. We find only one bound state, the $\frac{\text{3}}{\text{2}}^-$ ground state of \elem{Be}{9} with energies $-49.17\,\text{MeV}$ and $-54.87\,\text{MeV}$ for the NN+3N-induced and NN+3N-full Hamiltonians, respectively.

In Fig.~\ref{fig:Spectra_Nmax}(a) and (b) we show the energy spectrum of \elem{Be}{9} relative to the n-\elem{Be}{8} threshold computed with the NN+3N-full Hamiltonian for negative and positive parity, respectively. Each panel shows the convergence pattern of the energy levels with respect to the model-space size $N_{\text{max}}$ for the NCSM, i.e., without continuum degrees of freedom, and for the NCSMC, i.e., including continuum effects.

Comparing NCSM and NCSMC results for negative parity at fixed $N_{\text{max}}$, we find for all states significant contributions from the continuum coupling. The sole exception is the $\frac{\text{7}}{\text{2}}^-$ state, where the effects stay below 0.5\,MeV.
The NCSMC reduces the energy differences to the n-\elem{Be}{8} threshold compared to the NCSM for all states and for all $N_{\text{max}}$. Concerning the dependence on the model-space size for NCSMC increasing $N_{\text{max}}$  from 6 to 12 produces only small energy shifts, which are slightly larger for the higher-excited states but remain well below 0.5\,MeV. Hence, the NCSMC calculations are well converged, as already observed for the eigenphase shifts in Fig.~\ref{fig:NmaxDep}. This is different for the NCSM energies, which show significantly larger changes hinting at less converged calculations. This is of course expected, because all excited states of \elem{Be}{9} are resonances and the NCSM basis of $A$-body HO Slater determinants is not designed for a proper description of continuum states. Altogether, the NCSMC generally improves the agreement with experiment, and we find excellent agreement for the $\frac{\text{1}}{\text{2}}^-$ and second $\frac{\text{5}}{\text{2}}^-$ resonances beyond $N_{\text{max}}=10$. Note that also the \elem{Be}{9} ground-state energy is lowered by about 0.5\,MeV due the model-space extension by n-\elem{Be}{8} basis states and its agreement with experiment is improved.

\begin{figure}[t]
\includegraphics[width=0.87\columnwidth]{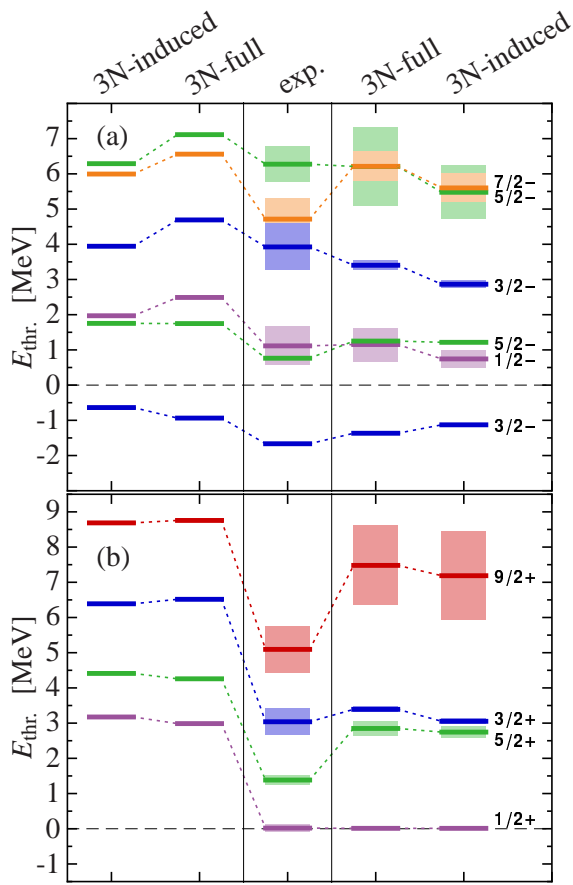}
\caption{(color online) Negative (a) and positive (b) parity spectrum of \elem{Be}{9} relative to the n-\elem{Be}{8} threshold at $N_{\text{max}} = 12$ and $11$, respectively. Shown are NCSM (first two columns) and NCSMC (last two columns) results compared to experiment~\cite{TiKe04}. First and last columns contain the energies for the NN+3N-induced and the second and fourth column for the NN+3N-full Hamiltonian, respectively. Shaded areas denote the width of the energy levels. Remaining parameters identical to Fig.~\ref{fig:Spectra_Nmax}.
}
\label{fig:Spectra}
\end{figure}

The most dramatic effects due to the continuum degrees of freedom are found in the positive-parity states, as evident from Fig.~\ref{fig:Spectra_Nmax}(b) by comparing the energies between the two approaches at fixed $N_{\text{max}}$. Once again, the NCSMC reduces all energy differences relative to the n-\elem{Be}{8} threshold compared to the NCSM, leading to an improved agreement with experiment. The agreement is particularly striking for the $S$-wave dominated $\frac{\text{1}}{\text{2}}^+$ state,
 whose energy at $N_{\text{max}}=7$ is shifted by the continuum degrees of freedom by about 5\,MeV right on top of its experimental position slightly above threshold, and remains practically constant when we increase the model-space size further to $N_{\text{max}}=11$. 
Also the remaining NCSMC energies are much less affected by increasing $N_{\text{max}}$ than the NCSM energies.
 We find the $\frac{\text{3}}{\text{2}}^+$ resonance, dominated by the $^4S_{\frac{3}{2}}$ partial wave, in good agreement with experiment, while discrepancies remain larger for the $\frac{\text{5}}{\text{2}}^+$ and $\frac{\text{9}}{\text{2}}^+$ resonances. Finally, we note that contributions from the broad $4^+$ state of \elem{Be}{8} might improve the description of the $\frac{\text{9}}{\text{2}}^+$ resonance of \elem{Be}{9}.

In Fig.~\ref{fig:Spectra} we study the effects of the initial chiral 3N interaction on the \elem{Be}{9} energy levels by comparing the spectrum to the one for the NN+3N-induced Hamiltonian including the SRG-induced 3N interactions only.
For negative parity, all states, except the first $\frac{\text{5}}{\text{2}}^-$ resonance, are sensitive to the inclusion of the initial chiral 3N interaction with effects of roughly similar size for both the NCSM and the NCSMC: the inclusion of the chiral 3N interaction increases the resonance energies relative to the threshold. Because the NCSM energy differences for the NN+3N-induced Hamiltonian are typically close to or above the experimental values, the agreement with experiment deteriorates when the initial 3N interaction is included. 
In contrast, the NCSMC energy differences for the NN+3N-induced Hamiltonian are typically below the experimental ones, and the overall agreement with experiment is clearly improved due to the initial chiral 3N interaction. 
This conclusion would have been opposite based on the NCSM alone, and highlights the importance of a proper treatment of the continuum as in the NCSMC to correctly assess the role of the 3N force. 
Interestingly, the $\frac{\text{5}}{\text{2}}^-$ resonance is not affected at all by the inclusion of the chiral 3N interactions, and for the $\frac{\text{7}}{\text{2}}^-$ state the energy shift caused by the chiral 3N interaction has the wrong sign, hinting at possible deficiencies in the spin-orbit structure of the initial Hamiltonian.
In Fig.~\ref{fig:Spectra}(b) we find the positive-parity states with both methods rather insensitive to initial 3N interactions. For the $\frac{9}{2}^+$ and $\frac{3}{2}^+$ states we find slightly increased energies and minor effects for the $\frac{5}{2}^+$ energy. The $\frac{\text{1}}{\text{2}}^{+}$ state is particularly unaffected and remains slightly above threshold in excellent agreement with experiment for both Hamiltonians. Note, however, that we found this state very weakly bound in calculations with NN forces only (not shown). 

Finally, for the NCSMC calculations in Fig.~\ref{fig:Spectra} we compare the extracted resonance widths to experiment. Overall the widths are of the same order of magnitude but typically smaller than the experimental ones, except for the $\frac{\text{9}}{\text{2}}^+$ and the $\frac{\text{5}}{\text{2}}^-$ resonances. For the $\frac{1}{2}^-$, $\frac{1}{2}^+$ and $\frac{5}{2}^+$ widths we find good agreement with experiment. In particular, the narrow $\frac{\text{5}}{\text{2}}^-$ resonance with experimental width of 0.78\,keV is also very narrow in our NCSMC calculations.
See Supplemental Material~\cite{Supp} for a table of the extracted energies and widths.


\paragraph{Conclusions.}
We have generalized the NCSMC approach to explicitly include 3N interactions with access to $p$- and lower-$sd$-shell target nuclei, and have studied the energy spectrum of \elem{Be}{9} as first application. We have found significant contributions of the continuum degrees of freedom, in particular for states with low angular momenta for which the centrifugal barrier is small or nonexistent. The continuum contributions significantly improve the model-space convergence, such that the \elem{Be}{9} spectrum is essentially converged at $N_{\text{max}}=6$. Furthermore, we have found the NCSMC particularly important for the assessment of the 3N interactions, which can be misleading based on NCSM calculations alone. With the NCSMC we found the chiral 3N interaction generally improving the agreement with experiment for the low-energy spectrum of \elem{Be}{9}. The sole exception is the $\frac{\text{7}}{\text{2}}^-$ state, which is rather insensitive to both additional continuum degrees of freedom and larger model spaces. Although we cannot rule out the relevance of cluster structures beyond the single-nucleon binary-cluster ansatz used here, one might expect larger sensitivities to the NCSM model-space size if such structures were to be relevant. Therefore, the present deviations from experiment are likely to be connected to deficiencies of the chiral NN+3N Hamiltonian.

Future work will use the $^9$Be wave function with proper asymptotic behaviour with respect to the n-$^8$Be threshold  to calculate various observables, including $E1$ transitions and the n-$^8$Be capture cross section. Furthermore, the formalism will be generalized to multi-nucleon projectiles, namely the deuteron, $^3$H, $^3$He and $^4$He.

\paragraph{Acknowledgments.}

Numerical calculations have been performed at the National Energy Research Scientific Computing Center (edison) supported by the Office of Science of the U.S. Department of Energy under Contract No. DE-AC02-05CH11231, at the LOEWE-CSC Frankfurt, and at the computing center of the TU Darmstadt (lichtenberg). Further, computing support for this work came in part from the LLNL institutional Computing Grand Challenge program and from an INCITE Award on the Titan supercomputer of the Oak Ridge Leadership Computing Facility (OLCF) at ORNL.

Supported by the Deutsche Forschungsgemeinschaft through contract SFB 634, by the Helmholtz International Center for FAIR (HIC for FAIR) within the LOEWE program of the State of Hesse, and the BMBF through contract 06DA7047I and from the NSERC grant No. 401945-2011. Prepared in part by LLNL under Contract DE-AC52-07NA27344 and supported by the U.S.~Department of Energy, Office of Science, Office of Nuclear Physics, Work Proposal Number SCW1158. TRIUMF receives funding via a contribution through the National Research Council Canada. 


\end{document}